# Topological Tunneling Magnetoresistance Driven by Type-II Weyl-Like States in the Room-Temperature Half-Metal Mn$_2$PC Monolayer

Wei Ma[#], Yu-Ting Wang[#], Wen-Bo Sun[#], Zhiheng Lv, Shuai Shi, Jian-Hong Rong[*], Tie-Lei Song[*], Zhi-Feng Liu[*]

*Inner Mongolia Key Laboratory of Microscale Physics and Atomic Manufacturing & Research Center for Quantum Physics and Technologies, School of Physical Science and Technology, Inner Mongolia University, Hohhot 010021, China*

**Abstract** We predict the tetragonal Mn$_2$PC monolayer to be a room-temperature ferromagnetic half-metal with a Curie temperature of 554 K. The spin-up channel hosts type-II Weyl-like crossings at the Fermi level with highly anisotropic band dispersion, whereas the spin-down channel is a wide-gap semiconductor. Topological edge states obtained from tight-binding calculations confirm the non-trivial bulk topology. Spin-orbit coupling opens a small gap of 11.2 meV at the Weyl-like crossings, generating pronounced Berry curvature and a sizable anomalous Hall conductivity near the Fermi level. Based on these properties, we propose topological tunneling magnetoresistance in a Mn$_2$PC-based magnetic tunnel junction: the parallel configuration conducts through fully spin-polarized Weyl-like carriers, while the antiparallel configuration is suppressed by the half-metallic gap, yielding a giant magnetoresistance ratio. The concurrent anomalous Hall effect in the conducting state provides an experimentally accessible signature of the topological carriers. These results identify the Mn$_2$PC monolayer as a promising platform for room-temperature topological spintronic devices.

[*]Corresponding authors: jhrong@imu.edu.cn; tlsong@imu.edu.cn; zfliu@imu.edu.cn

Magnetic tunnel junctions (MTJs), serving as the core architecture of modern spintronic technologies such as nonvolatile magnetic random-access memory (MRAM) [1], rely critically on the spin polarization of electrode materials. [2-3] In principle, half-metals with 100% spin polarization [4] provide ideal electrodes for achieving giant tunneling magnetoresistance (TMR). [5-6] In recent years, the discovery of intrinsic 2D ferromagnets, including $CrI_3$ [7], $Cr_2Ge_2Te_6$ [8], and $Fe_3GeTe_2$ [9], has created new opportunities for developing atomically thin, high-density spintronic devices. [10-13] Despite these advances, the practical development of 2D spintronic devices still faces several key challenges. First, although a few room-temperature 2D magnets have recently been predicted or experimentally stabilized through heterostructure engineering or external stimuli, [14-15] the Curie temperature ($T_C$) of most intrinsic 2D ferromagnets remains far below room temperature. [12-13] This limitation continues to hinder the realization of practical room-temperature spintronic applications. Second, the charge carriers in most proposed 2D half-metals exhibit conventional parabolic band dispersions. [16-21] Such trivial electronic states are sensitive to scattering from thermal fluctuations and lattice imperfections, degrading coherent spin transport in nanoscale devices. [22]

Introducing topological quantum states into spintronic systems offers a promising route to enhance carrier robustness and transport efficiency. [23-28] In particular, magnetic Weyl semimetals [29] host topologically protected band crossings accompanied by nontrivial Berry curvature and chiral quasiparticles, which can suppress certain backscattering processes. [27-28,30-32] Among them, type-II Weyl states are characterized by strongly tilted band dispersions [33-35], giving rise to pronounced anisotropy in carrier velocity near the Fermi surface. This anisotropic electronic structure is particularly suited for realizing directional quantum tunneling devices. [34-35] However, in most known magnetic Weyl materials, trivial bulk bands coexist at the Fermi level, [36-37] diluting the topological transport signatures and introducing leakage currents in the antiparallel MTJ configuration—thereby limiting the achievable magnetoresistance ratio. [35,38]

In this context, recent theoretical studies have proposed several 2D magnetic Weyl half-metal candidates, [24-26,39-44] with a few systems exhibiting type-II topological features. [45-47] Nevertheless, these works mainly focus on the static characterization of band topology. The integration of Weyl-related electronic properties into realistic spin-transport device concepts remains largely unexplored. In particular, achieving a single 2D material platform that simultaneously combines room-temperature ferromagnetism, intrinsic half-metallicity, and type-II Weyl-like topological states would provide a promising route for realizing high-contrast spintronic switching devices. In such a system, transport in an MTJ would be dominated by topological carriers within a single spin channel, while the opposite spin channel remains insulating due to the half-metallic gap.

Motivated by this possibility, we predict, based on first-principles calculations (see Sec. S1 in Supporting Information (SI) including Refs.[48-57]) and quantum transport simulations, that the tetragonal $Mn_2PC$ monolayer provides a promising platform for realizing the above combination. It exhibits robust room-temperature ferromagnetism with a Curie temperature of 554 K and a half-metallic electronic structure in which the spin-up channel hosts strongly anisotropic type-II Weyl-like states while the spin-down channel acts as a wide-gap semiconductor. Building on these features, we construct a homojunction MTJ model and propose the concept of topological tunneling magnetoresistance (TTMR). Transport calculations show that the transmission in the antiparallel configuration is strongly suppressed, enabling a high-contrast spin-switching behavior. In the parallel configuration, the tunneling transport is governed by the type-II Weyl-like carriers. Furthermore, the minute gap (11.2 meV) opened by spin-orbit coupling (SOC) transforms the linear crossings into massive topological states. This gap generates an intense Berry curvature hotspot near the Fermi level and leads to a sizable anomalous Hall response. The coexistence of efficient spin switching and a measurable anomalous Hall signal suggests a feasible route for integrating topological transport and spintronic functionality within a single 2D material platform.

**A. Crystal structure, stability, and synthesizability.** Our design of the 2D $Mn_2PC$ monolayer is motivated by the extensive experimental realization of layered transition metal pnictides, such as BaMn$_2$P$_2$ [58] and CaMn$_2$P$_2$. [59] These parent compounds are characterized by robust, covalently bonded [Mn$_2$P$_2$] layers that are intercalated by alkaline-earth metal atoms. Such quasi-2D architecture provides a natural structural blueprint for isolating the atomically thin Mn$_2$P$_2$ monolayer. [54] While previous theoretical studies have explored pristine Mn$_2$P$_2$ [54] and its Janus derivatives (e.g., Mn$_2$PSb [60] and Mn$_2$PAs [61]) as ferromagnetic (FM) half-metals, these systems intrinsically lack topological Weyl-like states, and their MTJ applications remain unexplored. To bridge this critical gap, a more pronounced structural and chemical asymmetry is required to effectively break the spatial inversion symmetry and induce non-trivial band topology. Driven by this deep theoretical motivation, we construct the asymmetric Mn$_2$PC monolayer by replacing one phosphorus layer in the Mn$_2$P$_2$ lattice with carbon atoms. The choice of carbon is physically justified by its smaller atomic radius and higher electronegativity compared to P, As, or Sb. Furthermore, this design is experimentally feasible, directly inspired by mature surface anion engineering techniques widely used to tailor transition metal phosphides under controlled environments. [62]

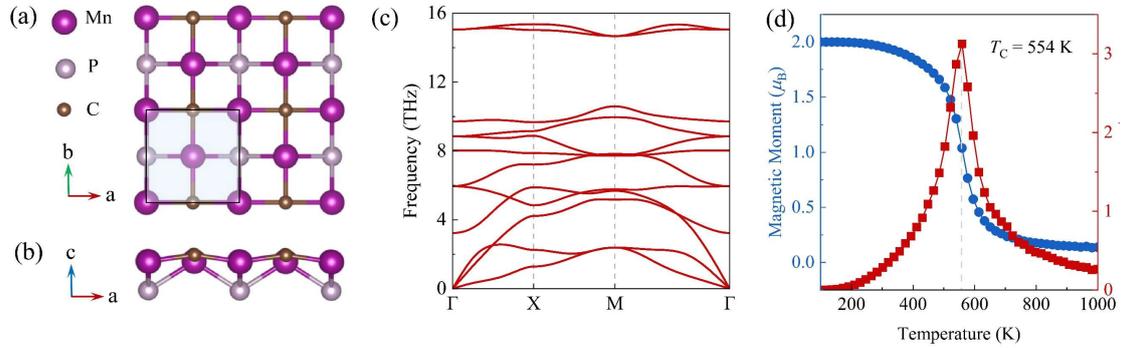

**FIG. 1.** (a) Top and (b) side views of the Janus Mn$_2$PC lattice; the shaded area indicates the primitive unit cell. (c) Phonon dispersion curves along the high-symmetry lines in the Brillouin zone. (d) Temperature-dependent magnetic moment (blue) and specific heat C$_v$ (red) from Monte Carlo simulations, indicating a Curie temperature of $T_C$=554 K. (e) Fluctuations of the potential energy (blue) and total magnetic moment (red) during AIMD simulations at 550 K. The inset displays the atomic structure snapshot after 5 ps.

To validate this theoretical design, we first relaxed the crystal structure using first-principles calculations. As shown in Figs. 1(a) and 1(b), $Mn_2PC$ crystallizes in a tetragonal lattice (space group *P4mm*, No. 99) with optimized in-plane lattice constants $a = b = 4.063$ Å. The central Mn layer is sandwiched between top P and bottom C layers, forming a Janus configuration with distinct bond lengths: Mn-P = 2.431 Å and Mn-C = 2.064 Å. This pronounced difference indicates stronger Mn-C bonding, yielding a formation energy that is 0.152 eV/atom lower than that of pristine $Mn_2P_2$. This energy advantage provides a strong thermodynamic driving force for experimental synthesis (e.g., via surface substitution or chemical vapor deposition). Crucially, this robust out-of-plane structural and chemical asymmetry explicitly breaks the spatial inversion symmetry—a fundamental prerequisite for generating non-trivial Berry curvature and the ensuing macroscopic topological transport.

Before exploring its topological properties, we first establish the stability of the $Mn_2PC$ monolayer. Dynamic stability at 0 K is confirmed by the absence of imaginary frequencies in the phonon dispersion [Fig. 1(c)]. Moreover, the calculated elastic constants ($C_{11} = 104$ N/m, $C_{12} = 89$ N/m, $C_{66} = 8$ N/m) satisfy the Born-Huang criteria ($C_{11} > 0$, $C_{66} > 0$, and $C_{11} > |C_{12}|$ [63]), confirming macroscopic mechanical stability. Regarding magnetic stability, total energy calculations confirm that the FM state remains the robust energetic ground state against various antiferromagnetic (AFM) configurations (Fig. S1 in Sec. S2 of SI). We then mapped the energy differences among these distinct magnetic configurations to extract the exchange coupling constants ($J_1 = 3.696$ meV, $J_2 = 2.097$ meV), thereby establishing a reliable spin Hamiltonian [see Fig. S1(a)]. The preferred magnetization direction is governed by the magnetic anisotropy energy (MAE), defined as the total energy difference between the in-plane and out-of-plane magnetization states (MAE = $E_x - E_z$). Our calculations yield a positive MAE of ~0.5 meV/unit cell, providing sufficient perpendicular magnetic anisotropy (PMA) with the *z*-axis as the intrinsic easy axis. This intrinsic out-of-plane anisotropy explicitly circumvents the Mermin-Wagner restriction [64], enabling the stabilization of true long-range 2D magnetic order at finite temperatures. Incorporating the exchange couplings and the PMA into an anisotropic Heisenberg model, our Monte Carlo (MC) simulations predict a remarkably high Curie temperature ($T_C$) of ~554 K [Fig. 1(d)]. For practical MTJ devices, such a robust out-of-plane MAE is highly advantageous: it provides a substantial energy barrier against thermal fluctuations, thereby ensuring robust non-volatility suitable for high-density spin-valve architectures.

Furthermore, spin-polarized *ab initio* molecular dynamics (AIMD) [65] simulations at 550 K (just below $T_C$) provide an important microscopic verification of the high-temperature structural integrity. Despite significant thermal fluctuations (Fig. S2 in SI), the structural framework remains fully intact over 5 ps without any bond breaking. The total magnetic moment also remains largely stable near its ground-state value (red line). This agreement between the dynamic AIMD simulations and the static MC model convincingly demonstrates the material's robust structural and perpendicular magnetic integrity even when approaching the phase transition point.

**B. Spin-polarized electronic structure and type-II Weyl-like states.** Having established the robust room-temperature FM ground state, we now investigate its spin-polarized electronic properties. Figs. 2(a) and 2(b) present the calculated band structure and the projected density of states at the PBE level without SOC. An intrinsic half-metallicity is immediately evident: the spin-up channel exhibits strongly dispersive metallic bands crossing the Fermi level ($E_F$), whereas the spin-down channel acts as a semiconductor with a wide bandgap of ~1.82 eV (3.75 eV in HSE06). Crucially, this half-metallic nature is further confirmed by HSE06 calculations [see Fig. S3(a) of Sec. S3 in SI]. The half-metallic gap ($E_{hg}$, defined as the minimum energy difference between the Fermi level and the spin-down band edges) is 0.50 eV at the PBE level (2.06 eV at the HSE06 level), which strongly suppresses thermally activated spin-flip excitations, effectively preserving full spin polarization at room temperature and laying a foundation for MTJ applications.

Focusing on the metallic spin-up channel, a prominent band crossing occurs along the high-symmetry X-M path [denoted as point D (0.5, 0.14797053, 0) in Fig. 2(c) and 2(d)], located slightly below the Fermi level at $E - E_F \sim -0.16$ eV. The orbital-projected band structure [Fig. 2(d)] reveals a clear band inversion driven by the P and C orbitals. Since the two intersecting bands belong to different irreducible representations ($\Gamma_1$ and

$\Gamma_2$) of the crystal's symmetry little group $C_{2v}$ along the X-M line, hybridization is forbidden, resulting in a symmetry-protected gapless nodal point in the absence of SOC. We note that while such isolated and fully spin-polarized linear band crossings in 2D systems are frequently referred to as "2D Weyl states/cones" in recent literature [43,46,66], rigorous definitions dictate that true Weyl nodes require a 3D momentum space to enclose a non-zero topological charge [67]. Therefore, we adopt the term "2D Weyl-like states/cones" to describe the spin-polarized linear band crossings in $Mn_2PC$ monolayer.

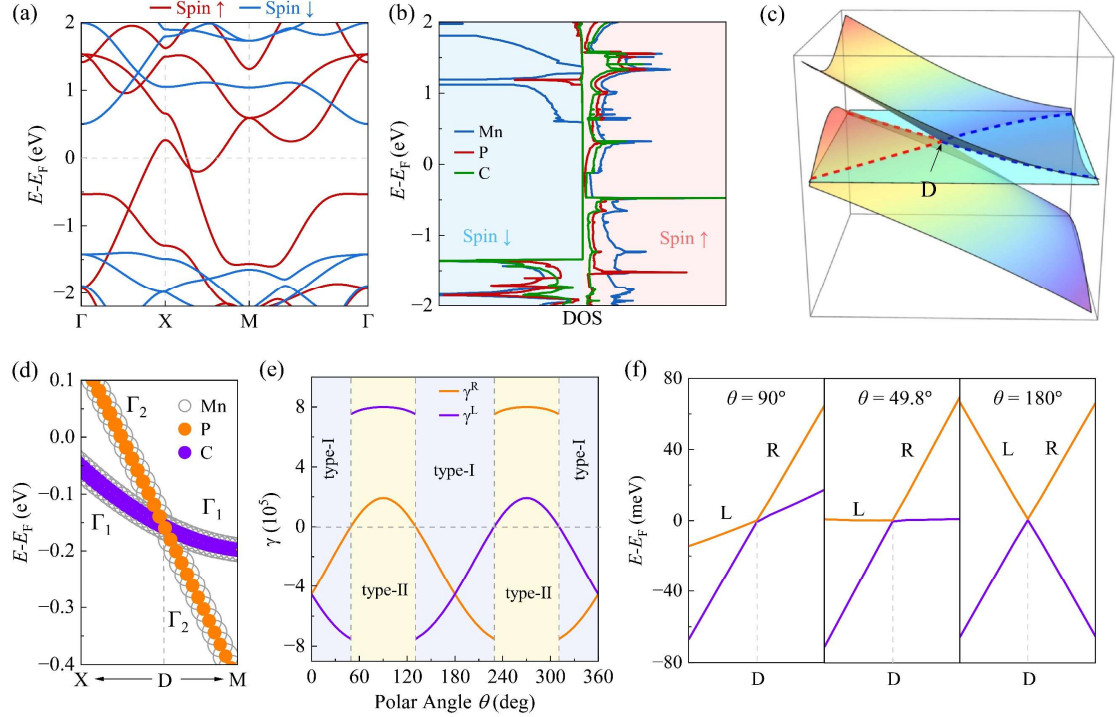

**FIG. 2.** (a) Spin-polarized band structure and (b) projected DOS of the $Mn_2PC$ monolayer without SOC. (c) 3D energy dispersion of the spin-up band crossing near the Fermi level around the D point. (d) Orbital-resolved band structure along the X-D-M for the spin-up channel. Labels $\Gamma_1/\Gamma_2$ denote irreducible representations. (e) The slope indices $\gamma^L$ and $\gamma^R$ as a function of the polar angle $\theta'$. (f) Band profiles at representative angles $\theta = 90°, 49.8°, 180°$.

Notably, the 3D band dispersion around point D [Fig. 2(c)] exhibits an overtilted cone profile, which is the hallmark of a type-II Weyl-like cone. [34] Consequently, the electron and hole pockets touch exactly at the nodal point. To quantitatively evaluate the anisotropic dispersion, we adopt the direction-dependent slope index $\gamma$, defined as $\gamma^{L(R)} = v_F^{L(R)} \text{sgn}(v_F^{R(L)})$ [68]. Here, $v_F^{L/R}$ represents the Fermi velocity of the left (L) or right (R) crossing band. By definition, a positive $\gamma$ indicates that the two bands share the same sign of slope, the essential criterion for a type-II Weyl-like cone. As illustrated in Figs. 2(e) and 2(f), the dispersion exhibits a dramatic transition: at $\theta' = 90°$, the slopes share the same sign, confirming the type-II nature; at $\theta' = 180°$, they adopt opposite signs (type-I-like). Most strikingly, at the critical angle $\theta' = 49.8°$, one band becomes completely flat ($v_F = 0$), while the other remains steeply dispersive. This extreme velocity anisotropy (ranging from 0 to $7.99 \times 10^5$ m/s, comparable to that reported in graphene $8.25 \times 10^5$ m/s [69]) not only provides a significant density-of-states enhancement associated with the flat-band pocket, but also offers a highly directional conducting channel for topological quantum tunneling.

**C. Topological edge states and SOC-driven massive anomalous Hall effect.** To evaluate the topological signatures of the spin-polarized type-II nodal points and their macroscopic transport consequences, we constructed a tight-binding (TB) Hamiltonian based on the maximally localized Wannier functions (MLWFs) [56]. As depicted in Fig. 3(a), the Wannier-interpolated bands well reproduce the DFT band structure for the spin-up channel, providing an accurate foundation for topological evaluations.

The fundamental hallmark of a 2D topological nodal point is the localized accumulation of Berry curvature. Without SOC, the Berry curvature distribution along the high-symmetry path [Fig. 3(b), purple line] exhibits a sharp and highly localized peak at point D, characteristic of a massless linear crossing. According to the bulk-boundary correspondence, this nontrivial topology manifests as localized edge states. Using the iterative Green's function method, we mapped the local density of states (LDOS) for a semi-infinite nanoribbon. In Fig. 3(c), distinct 1D topological edge states (colored red and blue for the left and right edges, respectively) traverse the bulk band projections, strongly confirming the nontrivial topology in the $Mn_2PC$ monolayer. Upon including SOC, the band degeneracy is lifted, opening a small gap of ~11.2 meV [Fig. S3(b) in SI] slightly below the Fermi level at $E − E_F \approx −0.162$ eV. This interaction transforms the massless crossing into a massive topological state. As evidenced by the orange line in Fig. 3(b), the SOC-induced mass term physically broadens the Berry curvature peak, distributing it over a finite momentum region around the avoided crossing.[70] Macroscopically, this localized Berry curvature hotspot acts as a strong fictitious magnetic field, driving transverse quantum transport. The intrinsic anomalous Hall conductivity (AHC, $\sigma_{xy}$) is obtained by integrating the Berry curvature over the Brillouin zone. Strikingly, rather than suppressing the topological signal, the SOC-induced gap activates a substantial enhancement of the AHC. As illustrated in Fig. 3(d), while the pristine non-SOC band structure yields a modest AHC peak of ~30 $(\Omega \cdot cm)^{-1}$, the inclusion of SOC dramatically amplifies the peak to ~230 $(\Omega \cdot cm)^{-1}$. The broadening and slight energy shift of this enhanced AHC peak reflect the integration of the dispersed Berry curvature across the strongly anisotropic type-II band dispersion, a distinctive transport footprint of massive topological carriers in magnetic systems.[28,71-72]

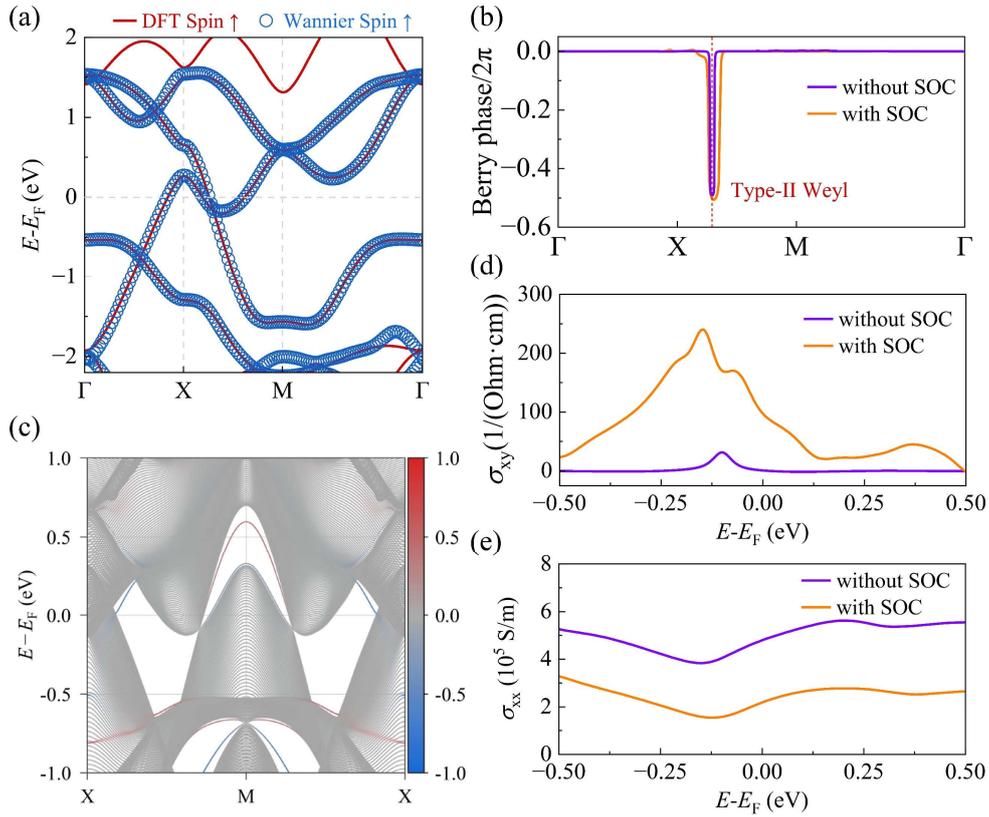

**FIG. 3.** (a) Comparison of electronic band structures from DFT (red solid lines) and the Wannier90 TB model (blue circles) for the spin-up channel. (b) Berry curvature along the high-symmetry $k$-path. (c) Projected 1D edge state spectrum of the $Mn_2PC$ monolayer. (d) Intrinsic anomalous Hall conductivity ($\sigma_{xy}$) and (e) longitudinal conductivity ($\sigma_{xx}$) as a function of the Fermi level shift.

Furthermore, it is essential to address the coexistence of this robust topological signal with the system's excellent metallic transport capabilities. Because the massive topological node is located at −0.162 eV, the actual Fermi level ($E_F = 0$) bypasses the SOC-induced gap and continues to intersect the strongly dispersive metallic bands. Consequently, even with the inclusion of SOC, the longitudinal conductivity ($\sigma_{xx}$) remains highly robust at ~$2.5 \times 10^5$ S/m in this energy regime [Fig. 3(e)]. The abundant conduction electrons at the Fermi surface are strongly scattered by the underlying topological Berry curvature hotspot, thereby continuously generating a macroscopic anomalous Hall signal. This unique energy alignment establishes the $Mn_2PC$ monolayer as an attractive platform for exploring topology-driven spintronic phenomena at practical temperatures.

**D. Topological Tunneling Magnetoresistance and Ultra-High Contrast Spin Switch.** To harness these extraordinary properties for practical applications, we propose a 2D homojunction MTJ device. As schematically illustrated in Fig. 4(a), the MTJ consists of two identical $Mn_2PC$ monolayers acting as ferromagnetic electrodes, separated by a generic non-magnetic insulating barrier layer (IL), which is modeled as a tunable tunneling barrier within the TB framework. The quantum transport is governed by the relative magnetization orientation of the two electrodes.

To quantitatively evaluate the device performance, we constructed an effective TB Hamiltonian capturing the massive type-II Weyl-like physics around the Fermi level (see Sec. S4 in the SI). The transmission probability was computed using the Landauer-Büttiker formalism [73-74] as implemented in the Kwant code. [75] Fig. 4(b) presents the energy-dependent transmission spectra for both the parallel (P) and antiparallel (AP) configurations, varying the barrier length $N_x$ from 1 to 5 unit cells while maintaining a fixed transverse device width of $N_y = 20$ unit cells. In the P configuration (ON-state), the spin-up conducting channels of both electrodes are closely aligned, and the tunneling current is carried by highly conductive type-II Weyl-like carriers. As dictated by quantum tunneling principles, the overall transmission probability gradually decays as the barrier length $N_x$ increases; nevertheless, it remains highly robust across the entire energy window, confirming the excellent transport efficiency of the topological channels. Conversely, in the AP configuration (OFF-state), the spin-up Weyl-like carriers injected from the left electrode encounter the massive insulating gap of 1.82 eV (3.75 eV, HSE06) of the right electrode's spin-down channel. Within the spin-conserving coherent transport picture, and given the absence of available states within the energy gap, the tunneling pathway is entirely blocked [Fig. 4(a)]. Consequently, as shown by the red line in Fig. 4(b), the transmission probability in the AP state is nearly zero ($T_{AP} \approx 0$) within our coherent transport framework.

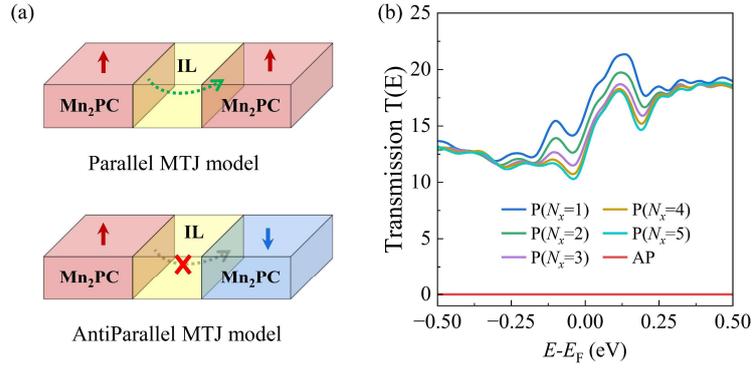

**FIG. 4.** (a) Schematic of the 2D $Mn_2PC$-based homojunction MTJ in parallel (top) and antiparallel (bottom) configurations. Green arrows and red crosses indicate allowed and blocked tunneling pathways, respectively. (b) Calculated energy-dependent transmission spectra for parallel and antiparallel MTJ models with different barrier lengths.

While practical devices at finite temperatures may exhibit a minute non-zero $T_{AP}$ due to defect-assisted tunneling or thermal spin-flip scattering, this ideal mismatch leads to an ultra-high contrast turn-off. Following the standard definition of the tunneling magnetoresistance ratio [5], TMR = $(T_P - T_{AP})/T_{AP}$, the vanishing denominator results in an extremely large magnetoresistance ratio. Recently, the concept of exploiting topological band structures to enhance tunneling magnetoresistance has garnered significant theoretical interest. For instance, gigantic TMR driven by chirality-magnetization locking has been proposed in 3D magnetic Weyl semimetal junctions. [76] However, our 2D $Mn_2PC$ MTJ operates on a distinct and highly practical physical paradigm: the OFF-state is rigorously blockaded by the massive half-metallic gap, while the ON-state transmission is governed by the momentum-filtering effect of highly directional massive Weyl-like carriers. More importantly, unlike conventional half-metallic MTJs or previously proposed 3D theoretical models, the ON-state conduction here is dominated by topologically non-trivial massive carriers, which simultaneously generate a substantial AHC. Therefore, we refer to this mechanism as topological tunneling magnetoresistance (TTMR). The realization of TTMR not only guarantees an ultimate ON/OFF switch ratio for high-density magnetic memory but also allows the macroscopic read-out of the underlying topological nature through the concurrent large AHC in the ON-state. This establishes the $Mn_2PC$ MTJ as a promising platform for next-generation topological spintronics.

In summary, we propose the 2D Janus $Mn_2PC$ monolayer as a robust material platform for room-temperature topological spintronics. Our comprehensive theoretical investigation reveals the following key findings: (i) the material exhibits excellent dynamic, mechanical, and thermal stability, featuring a high Curie temperature (~554 K) driven by strong exchange interactions; (ii) it possesses an intrinsic half-metallic electronic structure, where the spin-down channel acts as a wide-gap semiconductor and the spin-up channel hosts strongly anisotropic type-II Weyl-like states. Crucially, spin-orbit coupling transforms these linear crossings into massive topological states, thereby generating a large anomalous Hall conductivity; and (iii) the proposed homojunction MTJ demonstrates topological tunneling magnetoresistance. Enabled by the large perpendicular magnetic anisotropy and the massive half-metallic gap, this device achieves an ultra-high contrast spin turn-off in the antiparallel state. Its ON-state is governed by the momentum-filtering effect of highly directional massive Weyl-like carriers, yielding a very large magnetoresistance ratio and allowing concurrent state read-out via the substantial anomalous Hall conductivity. Our findings not only overcome the low-temperature and trivial-carrier limitations of conventional 2D spintronic systems, but also provide a practical theoretical framework for next-generation ultra-low-power topological spintronic devices operating at room temperature.

## Acknowledgments

This work is supported by National Natural Science Foundation of China (12464040, 11964023), Natural Science Foundation of Inner Mongolia Autonomous Region (2021JQ-001), and the 2020 Institutional Support Program for Youth Science and Technology Talents in Inner Mongolia Autonomous Region (NJYT-20-B02).

## References

[1] B. Dieny, I. L. Prejbeanu, K. Garello, P. Gambardella, P. Freitas, R. Lehndorff et al., "Opportunities and challenges for spintronics in the microelectronics industry," *Nat. Electron.* **3** (8), 446-459 (2020).

[2] W. H. Butler, X. G. Zhang, T. C. Schulthess, and J. M. MacLaren, "Spin-dependent tunneling conductance of Fe/MgO/Fe sandwiches," *Phys. Rev. B* **63** (5), 054416 (2001).

[3] S. Yuasa, T. Nagahama, A. Fukushima, Y. Suzuki, and K. Ando, "Giant room-temperature magnetoresistance in single-crystal Fe/MgO/Fe magnetic tunnel junctions," *Nat. Mater.* **3** (12), 868-871 (2004).

[4] R. A. de Groot, F. M. Mueller, P. G. v. Engen, and K. H. J. Buschow, "New class of materials: Half-metallic ferromagnets," *Phys. Rev. Lett.* **50** (25), 2024-2027 (1983).

[5] M. Julliere, "Tunneling between ferromagnetic films," *Phys. Lett. A* **54** (3), 225-226 (1975).

[6] M. I. Katsnelson, V. Y. Irkhin, L. Chioncel, A. I. Lichtenstein, and R. A. de Groot, "Half-metallic ferromagnets: From band structure to many-body effects," *Rev. Mod. Phys.* **80** (2), 315-378 (2008).

[7] B. Huang, G. Clark, E. Navarro-Moratalla, D. R. Klein, R. Cheng, K. L. Seyler et al., "Layer-dependent ferromagnetism in a van der Waals crystal down to the monolayer limit," *Nature* **546** (7657), 270-273 (2017).

[8] C. Gong, L. Li, Z. Li, H. Ji, A. Stern, Y. Xia et al., "Discovery of intrinsic ferromagnetism in two-dimensional van der Waals crystals," *Nature* **546** (7657), 265-269 (2017).

[9] Y. Deng, Y. Yu, Y. Song, J. Zhang, N. Z. Wang, Z. Sun et al., "Gate-tunable room-temperature ferromagnetism in two-dimensional $Fe_3GeTe_2$," *Nature* **563** (7729), 94-99 (2018).

[10] C. Gong, and X. Zhang, "Two-dimensional magnetic crystals and emergent heterostructure devices," *Science* **363** (6428), eaav4450 (2019).

[11] Z. Jia, M. Zhao, Q. Chen, Y. Tian, L. Liu, F. Zhang et al., "Spintronic devices upon 2D magnetic materials and heterojunctions," *ACS Nano* **19** (10), 9452-9483 (2025).

[12] X. Jiang, Q. Liu, J. Xing, N. Liu, Y. Guo, Z. Liu et al., "Recent progress on 2D magnets: Fundamental mechanism, structural design and modification," *Appl. Phys. Rev.* **8** (3), (2021).

[13] M. Gibertini, M. Koperski, A. F. Morpurgo, and K. S. Novoselov, "Magnetic 2D materials and heterostructures," *Nat. Nanotechnol.* **14** (5), 408-419 (2019).

[14] G. Zhang, F. Guo, H. Wu, X. Wen, L. Yang, W. Jin et al., "Above-room-temperature strong intrinsic ferromagnetism in 2D van der Waals $Fe_3GaTe_2$ with large perpendicular magnetic anisotropy," *Nat. Commun.* **13** (1), 5067 (2022).

[15] X. Zhang, Q. Lu, W. Liu, W. Niu, J. Sun, J. Cook et al., "Room-temperature intrinsic ferromagnetism in epitaxial $CrTe_2$ ultrathin films," *Nat. Commun.* **12** (1), 2492 (2021).

[16] M. Ashton, D. Gluhovic, S. B. Sinnott, J. Guo, D. A. Stewart, and R. G. Hennig, "Two-dimensional intrinsic half-metals with large spin gaps," *Nano Lett.* **17** (9), 5251-5257 (2017).

[17]S.-J. Gong, C. Gong, Y.-Y. Sun, W.-Y. Tong, C.-G. Duan, J.-H. Chu et al., "Electrically induced 2D half-metallic antiferromagnets and spin field effect transistors," *Proc. Natl. Acad. Sci. U.S.A.* **115** (34), 8511-8516 (2018).

[18]X. Li, and J. Yang, "First-principles design of spintronics materials," *Natl. Sci. Rev.* **3** (3), 365-381 (2016).

[19]J. Liu, Z. Liu, T. Song, and X. Cui, "Computational search for two-dimensional intrinsic half-metals in transition-metal dinitrides," *J. Mater. Chem. C* **5** (3), 727-732 (2017).

[20]R.-Z. Zhang, Y.-Y. Zhang, and S. Du, "Design of two-dimensional half-metals with large spin gaps," *Phys. Rev. B* **110** (8), 085130 (2024).

[21]X. Li, and J. Yang, "Low-dimensional half-metallic materials: Theoretical simulations and design," *Wiley Interdiscip. Rev.:Comput. Mol. Sci.* **7** (4), e1314 (2017).

[22]X. Lin, W. Yang, K. L. Wang, and W. Zhao, "Two-dimensional spintronics for low-power electronics," *Nat. Electron.* **2** (7), 274-283 (2019).

[23]H. Ishizuka, and Y. Motome, "Dirac half-metal in a triangular ferrimagnet," *Phys. Rev. Lett.* **109** (23), 237207 (2012).

[24]Z. Liu, J. Liu, and J. Zhao, "$YN_2$ monolayer: Novel p-state Dirac half metal for high-speed spintronics," *Nano Res.* **10** (6), 1972-1979 (2017).

[25]Q. Sun, and N. Kioussis, "Prediction of manganese trihalides as two-dimensional Dirac half-metals," *Phys. Rev. B* **97** (9), 094408 (2018).

[26]X. Wang, T. Li, Z. Cheng, X.-L. Wang, and H. Chen, "Recent advances in Dirac spin-gapless semiconductors," *Appl. Phys. Rev.* **5** (4), (2018).

[27]L. Šmejkal, Y. Mokrousov, B. Yan, and A. H. MacDonald, "Topological antiferromagnetic spintronics," *Nat. Phys.* **14** (3), 242-251 (2018).

[28]E. Liu, Y. Sun, N. Kumar, L. Muechler, A. Sun, L. Jiao et al., "Giant anomalous Hall effect in a ferromagnetic kagome-lattice semimetal," *Nat. Phys.* **14** (11), 1125-1131 (2018).

[29]X. Wan, A. M. Turner, A. Vishwanath, and S. Y. Savrasov, "Topological semimetal and Fermi-arc surface states in the electronic structure of pyrochlore iridates," *Phys. Rev. B* **83** (20), 205101 (2011).

[30]M. M. Piva, J. C. Souza, V. Brousseau-Couture, S. Sorn, K. R. Pakuszewski, J. K. John et al., "Topological features in the ferromagnetic Weyl semimetal CeAlSi: Role of domain walls," *Phys. Rev. Res.* **5** (1), 013068 (2023).

[31]O. K. Forslund, X. Liu, S. Shin, C. Lin, M. Horio, Q. Wang et al., "Anomalous Hall effect due to magnetic fluctuations in a ferromagnetic Weyl semimetal," *Phys. Rev. Lett.* **134** (12), 126602 (2025).

[32]I. Belopolski, R. Watanabe, Y. Sato, R. Yoshimi, M. Kawamura, S. Nagahama et al., "Synthesis of a semimetallic Weyl ferromagnet with point Fermi surface," *Nature* **637** (8048), 1078-1083 (2025).

[33]P. Li, Y. Wen, X. He, Q. Zhang, C. Xia, Z.-M. Yu et al., "Evidence for topological type-II Weyl semimetal $WTe_2$," *Nat. Commun.* **8** (1), 2150 (2017).

[34]A. A. Soluyanov, D. Gresch, Z. Wang, Q. Wu, M. Troyer, X. Dai et al., "Type-II Weyl semimetals," *Nature* **527** (7579), 495-498 (2015).

[35]B. Yan, and C. Felser, "Topological materials: Weyl semimetals," *Annu. Rev. Condens. Matter Phys.* **8**, 337-354 (2017).

[36]I. Belopolski, K. Manna, D. S. Sanchez, G. Chang, B. Ernst, J. Yin et al., "Discovery of topological Weyl fermion lines and drumhead surface states in a room temperature magnet," *Science* **365** (6459), 1278-1281 (2019).

[37]J. Noky, J. Gooth, C. Felser, and Y. Sun, "Characterization of topological band structures away from

the Fermi level by the anomalous Nernst effect," *Phys. Rev. B* **98** (24), 241106 (2018).

[38] Z. He, C.-T. Chou, E. Park, A. C. Foucher, B. C. McGoldrick, Q. Wang et al., "Magnetic tunnel junctions featuring the topological Weyl semimetal Co$_2$MnGa," *Phys. Rev. Appl.* **22** (4), 044024 (2024).

[39] H. Huan, Y. Xue, B. Zhao, H. Bao, L. Liu, and Z. Yang, "Tunable Weyl half-semimetals in two-dimensional iron-based materials *M*FeSe (*M*=Tl,In,Ga)," *Phys. Rev. B* **106** (12), 125404 (2022).

[40] Q. Li, L. Chen, R.-W. Zhang, and B. Fu, "Type-III Weyl semi-half-metal in an ultralight monolayer Li$_2$N," *Phys. Rev. B* **112** (19), 195126 (2025).

[41] W. Meng, X. Zhang, W. Yu, Y. Liu, L. Tian, X. Dai et al., "Multiple Weyl fermions and tunable quantum anomalous Hall effect in 2D half-metal with huge spin-related energy gap," *Appl. Surf. Sci.* **551**, 149390 (2021).

[42] W. Xu, J. Yi, H. Huan, B. Zhao, Y. Xue, and Z. Yang, "Two-dimensional half Chern-Weyl semimetal with multiple screw axes," *Phys. Rev. B* **106** (20), 205108 (2022).

[43] J.-Y. You, C. Chen, Z. Zhang, X.-L. Sheng, S. A. Yang, and G. Su, "Two-dimensional Weyl half-semimetal and tunable quantum anomalous Hall effect," *Phys. Rev. B* **100** (6), 064408 (2019).

[44] B. Zhang, X. Chen, and J. Wang, "Multiple Dirac states in two-dimensional topological half-metallic CrB$_2$C$_2$," *Appl. Phys. Lett.* **119** (16), (2021).

[45] W. Meng, X. Zhang, Y. Liu, L. Wang, X. Dai, and G. Liu, "Two-dimensional Weyl semimetal with coexisting fully spin-polarized type-I and type-II Weyl points," *Appl. Surf. Sci.* **540**, 148318 (2021).

[46] T. He, X. Zhang, Y. Liu, X. Dai, G. Liu, Z.-M. Yu et al., "Ferromagnetic hybrid nodal loop and switchable type-I and type-II Weyl fermions in two dimensions," *Phys. Rev. B* **102** (7), 075133 (2020).

[47] Y. Shi, L. Li, X. Cui, T. Song, and Z. Liu, "MnNBr monolayer: A high-temperature ferromagnetic half-metal with type-II Weyl fermions," *Phys. Status Solidi RRL* **15** (7), 2100115 (2021).

[48] G. Kresse, and J. Furthmüller, "Efficient iterative schemes for ab initio total-energy calculations using a plane-wave basis set," *Phys. Rev. B* **54** (16), 11169-11186 (1996).

[49] J. P. Perdew, K. Burke, and M. Ernzerhof, "Generalized gradient approximation made simple," *Phys. Rev. Lett.* **77** (18), 3865-3868 (1996).

[50] G. Kresse, and D. Joubert, "From ultrasoft pseudopotentials to the projector augmented-wave method," *Phys. Rev. B* **59** (3), 1758-1775 (1999).

[51] H. J. Monkhorst, and J. D. Pack, "Special points for Brillouin-zone integrations," *Phys. Rev. B* **13** (12), 5188-5192 (1976).

[52] A. Togo, and I. Tanaka, "First principles phonon calculations in materials science," *Scr. Mater.* **108**, 1-5 (2015).

[53] L. Wang, T. Maxisch, and G. Ceder, "Oxidation energies of transition metal oxides within the GGA+U framework," *Phys. Rev. B* **73** (19), 195107 (2006).

[54] B. Wang, Y. Zhang, L. Ma, Q. Wu, Y. Guo, X. Zhang et al., "MnX (X = P, As) monolayers: A new type of two-dimensional intrinsic room temperature ferromagnetic half-metallic material with large magnetic anisotropy," *Nanoscale* **11** (10), 4204-4209 (2019).

[55] J. Heyd, G. E. Scuseria, and M. Ernzerhof, "Hybrid functionals based on a screened Coulomb potential," *J. Chem. Phys.* **118**, 8207-8215 (2003).

[56] I. Souza, N. Marzari, and D. Vanderbilt, "Maximally localized Wannier functions for entangled energy bands," *Phys. Rev. B* **65** (3), 035109 (2001).

[57] G. Pizzi, D. Volja, B. Kozinsky, M. Fornari, and N. Marzari, "BoltzWann: A code for the evaluation of thermoelectric and electronic transport properties with a maximally-localized Wannier functions basis," *Comput. Phys. Commun.* **185** (1), 422-429 (2014).

[58]S. W. Lovesey, and D. D. Khalyavin, "Dirac multipoles in diffraction by the layered room-temperature antiferromagnets BaMn$_2$P$_2$ and BaMn$_2$As$_2$," *Phys. Rev. B* **98** (5), 054434 (2018).

[59]J. Zeng, S. Qin, C. Le, and J. Hu, "Magnetism and superconductivity in the layered hexagonal transition metal pnictides," *Phys. Rev. B* **96** (17), 174506 (2017).

[60]Q. Ma, W. Wan, Y. Li, and Y. Liu, "First principles study of 2D half-metallic ferromagnetism in Janus Mn$_2$XSb (X = As, P) monolayers," *Appl. Phys. Lett.* **120** (11), 112402 (2022).

[61]H. Zeng, S. Jin, J. Wang, Y. Hu, and X. Fan, "Ferromagnetic half-metal with high Curie temperature: Janus Mn2PAs monolayer," *J. Mater. Sci.* **56** (23), 13215-13226 (2021).

[62]Y. Shi, and B. Zhang, "Recent advances in transition metal phosphide nanomaterials: Synthesis and applications in hydrogen evolution reaction," *Chem. Soc. Rev.* **45** (6), 1529-1541 (2016).

[63]F. Mouhat, and F.-X. Coudert, "Necessary and sufficient elastic stability conditions in various crystal systems," *Phys. Rev. B* **90** (22), 224104 (2014).

[64]N. D. Mermin, and H. Wagner, "Absence of ferromagnetism or antiferromagnetism in one-or two-dimensional isotropic Heisenberg models," *Phys. Rev. Lett.* **17** (22), 1133 (1966).

[65]G. Kresse, and J. Hafner, "Ab initio molecular dynamics for liquid metals," *Phys. Rev. B* **47** (1), 558 (1993).

[66]Q. Lu, P. S. Reddy, H. Jeon, A. R. Mazza, M. Brahlek, W. Wu et al., "Realization of a two-dimensional Weyl semimetal and topological Fermi strings," *Nat. Commun.* **15** (1), 6001 (2024).

[67]N. P. Armitage, E. J. Mele, and A. Vishwanath, "Weyl and Dirac semimetals in three-dimensional solids," *Rev. Mod. Phys.* **90** (1), 015001 (2018).

[68]L. Cui, T. Song, J. Cai, X. Cui, Z. Liu, and J. Zhao, "Three-dimensional borophene: A light-element topological nodal-line semimetal with direction-dependent type-II Weyl fermions," *Phys. Rev. B* **102** (15), 155133 (2020).

[69]Z. Wang, X.-F. Zhou, X. Zhang, Q. Zhu, H. Dong, M. Zhao et al., "Phagraphene: A Low-Energy Graphene Allotrope Composed of 5-6-7 Carbon Rings with Distorted Dirac Cones," *Nano Lett.* **15** (9), 6182-6186 (2015).

[70]D. Xiao, M.-C. Chang, and Q. Niu, "Berry phase effects on electronic properties," *Rev. Mod. Phys.* **82** (3), 1959-2007 (2010).

[71]L. Ye, M. Kang, J. Liu, F. Von Cube, C. R. Wicker, T. Suzuki et al., "Massive Dirac fermions in a ferromagnetic kagome metal," *Nature* **555** (7698), 638-642 (2018).

[72]K. Kim, J. Seo, E. Lee, K.-T. Ko, B. Kim, B. G. Jang et al., "Large anomalous Hall current induced by topological nodal lines in a ferromagnetic van der Waals semimetal," *Nat. Mater.* **17** (9), 794-799 (2018).

[73]R. Landauer, "Electrical resistance of disordered one-dimensional lattices," *Philos. Mag.* **21** (172), 863-867 (1970).

[74]M. Büttiker, "Four-terminal phase-coherent conductance," *Phys. Rev. Lett.* **57** (14), 1761 (1986).

[75]C. W. Groth, M. Wimmer, A. R. Akhmerov, and X. Waintal, "Kwant: A software package for quantum transport," *New J. Phys.* **16** (6), 063065 (2014).

[76]D. De Sousa, C. Ascencio, P. M. Haney, J.-P. Wang, and T. Low, "Gigantic tunneling magnetoresistance in magnetic Weyl semimetal tunnel junctions," *Phys. Rev. B* **104** (4), L041401 (2021).